\documentclass{article}
\usepackage{spconf,amsmath,graphicx}
\usepackage{booktabs}
\usepackage{hyperref}
\usepackage{multirow}
\usepackage{color}
\usepackage{subfigure}

\title{Soft label coding for end-to-end sound source localization with ad-hoc microphone arrays}
\name{Linfeng Feng, Yijun Gong, Xiao-Lei Zhang}
\address{{School of Marine Science and Technology, Northwestern Polytechnical University, China}\\
    {Research \& Development Institute of Northwestern Polytechnical University in Shenzhen, China}\\
    fenglinfeng@mail.nwpu.edu.cn, gongyj@mail.nwpu.edu.cn, xiaolei.zhang@nwpu.edu.cn
   \thanks{Xiao-Lei Zhang is the corresponding author.}
   \thanks{This work was supported in
part by the Project of the Science, Technology, and Innovation
Commission of Shenzhen Municipality, China under Grant JSGG20210802152546026 and 
JCYJ20210324143006016, and in part by the National Science
Foundation of China (NSFC) under Grant 62176211.}
    }

\begin{document}
%
\maketitle
\begin{abstract}
Recently, an end-to-end two-dimensional sound source localization algorithm with ad-hoc microphone arrays formulates the sound source localization problem as a classification problem. The algorithm divides the target indoor space into a set of local areas, and predicts the local area where the speaker locates. However, the local areas are encoded by one-hot code, which may lose the connections between the local areas due to quantization errors. In this paper, we propose a new soft label coding method, named \textit{label smoothing}, for the classification-based two-dimensional sound source location with ad-hoc microphone arrays.
The core idea is to take the geometric connection between the classes into the label coding process.The first one is named static soft label coding (SSLC), which modifies the one-hot codes into soft codes based on the distances between the local areas. Because SSLC is handcrafted which may not be optimal, the second one, named dynamic soft label coding (DSLC), further rectifies SSLC, by learning the soft codes according to the statistics of the predictions produced by the classification-based localization model in the training stage. Experimental results show that the proposed methods can effectively improve the localization accuracy.
\end{abstract}
\section{Introduction}
\label{sec:intro}

Sound source localization is a problem of estimating the relative positions of sound sources to the locations of microphone arrays.
It has a wide range of applications, such as speech enhancement and separation \cite{wang2018supervised}, speech recognition \cite{spille2018comparing}, hearing aids \cite{van2011sound}, etc. Conventionally, the problem is simplified to the estimation of the direction of arrival (DOA) of sound sources. Representative approaches include time difference of arrival (TDOA) \cite{knapp1976generalized}, steered response power with phase transform (SRP-PHAT) \cite{dibiase2001robust}, and multiple signal classification (MUSIC)\cite{schmidt1986multiple}.
Recently, there is increased research interest in using deep neural network (DNN) for localizing sound sources. It takes spatial features, e.g. phase spectrogram \cite{chakrabarty2019multi}, cross-correlation \cite{xiao2015learning}, or spatial pseudo-spectra \cite{nguyen2020robust}, as the input of DNN to estimate the DOA. The aforementioned methods are based on a single microphone array.

If we jointly use multiple microphone arrays for sound source localization, then we might get the two-dimensional (2D) localization of sound sources directly. Ad-hoc microphone array, which collaboratively organizes a set of randomly distributed microphone arrays in space, is a solution to the problem. Early methods are mostly based conventional signal processing methods \cite{griffin2015localizing}. Recently, deep-learning-based methods were studied \cite{hahmann2022sound}, which is the focus of the paper. Although the pioneering works in \cite{vesperini2018localizing} made significant contribution to the direction, they make strong assumptions to the ad-hoc microphone arrays, such as fixed positions of microphone nodes at both the training and test stages, which do not fully utilize the advantage of the flexibility of ad-hoc arrays. Although the method in \cite{liu2022deep} can handle flexible ad-hoc arrays that consists of any number of randomly distributed ad-hoc nodes, it is a stage-wise method. Its deep models are used only at each ad-hoc node, leaving the 2-dimensional localization process an independent geometric approach.

As we know, a key advantage of deep learning is that it can address a task in an end-to-end manner. In \cite{gong2022e2e}, an end-to-end 2D sound source localization model with ad-hoc arrays was proposed. It formulates the 2D sound source localization problem as a classification problem, where each class represents a local area of a targeted room and is encoded by a one-hot code. Moreover, the method is effective in the situation where each ad-hoc node contains only a single microphone. However, the position of a speaker is estimated as a one-hot code, which can be problematic. For one-hot, the correct class is the same distance from any incorrect class.

To address the problem, we propose a label smoothing algorithm to remedy the weakness of the one-hot codes.
Specifically, we first propose static soft label coding (SSLC) based on the distances between the centers of local regions instead of the one-hot codes. Because SSLC is handcrafted that may not be accurate enough, we further propose to remedy SSLC by dynamic soft label coding (DSLC). DSLC is learned from the inference statistics produced by the localization model at the training stage. Experimental results demonstrate the effectiveness of the proposed method.

\section{Related work}

The proposed method is related to the weakness of one-hot codes. Specifically, one-hot codes set the distance between any pair of classes the same, and do not consider the inter-class similarity within classes. To address the problem, \textit{label smoothing} was originally proposed for computer vision \cite{szegedy2016rethinking}, which improves the generalization performance of classification models. However, it still ignores the inter-class similarity. An online label smoothing method based on the statistics of the predictions was proposed in \cite{zhang2021delving}. The generated soft labels reflect the inter-class similarity. However, directly applying the above methods to sound source localization makes the classification model non-convergence and yields poor performance. In this paper, we propose to jointly optimize the above two soft labels for the weakness of the one-hot coding in the end-to-end sound source localization.

%

\begin{figure}[t]
    \centering
    \includegraphics[width=0.22\textwidth]{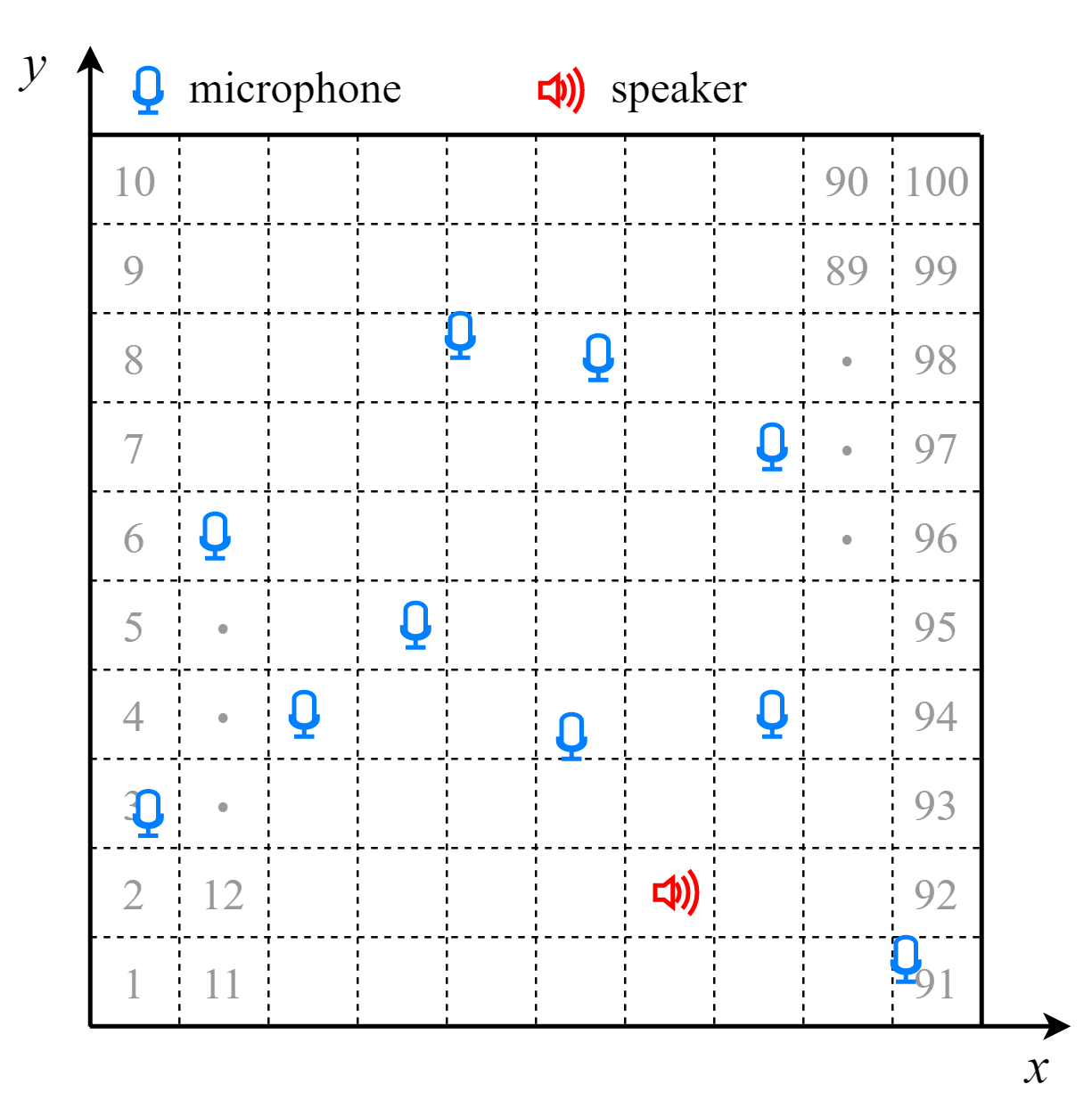}
    \caption{The classification-based formulation of 2D sound source location with ad-hoc microphone arrays, where a room is partitioned into multiple local areas. }
    \label{fig:room}
\end{figure}

\section{Method}
\label{sec:method}

\subsection{End-to-end sound source localization}

\begin{figure}[t]
    \centering
    \includegraphics[width=0.45\textwidth]{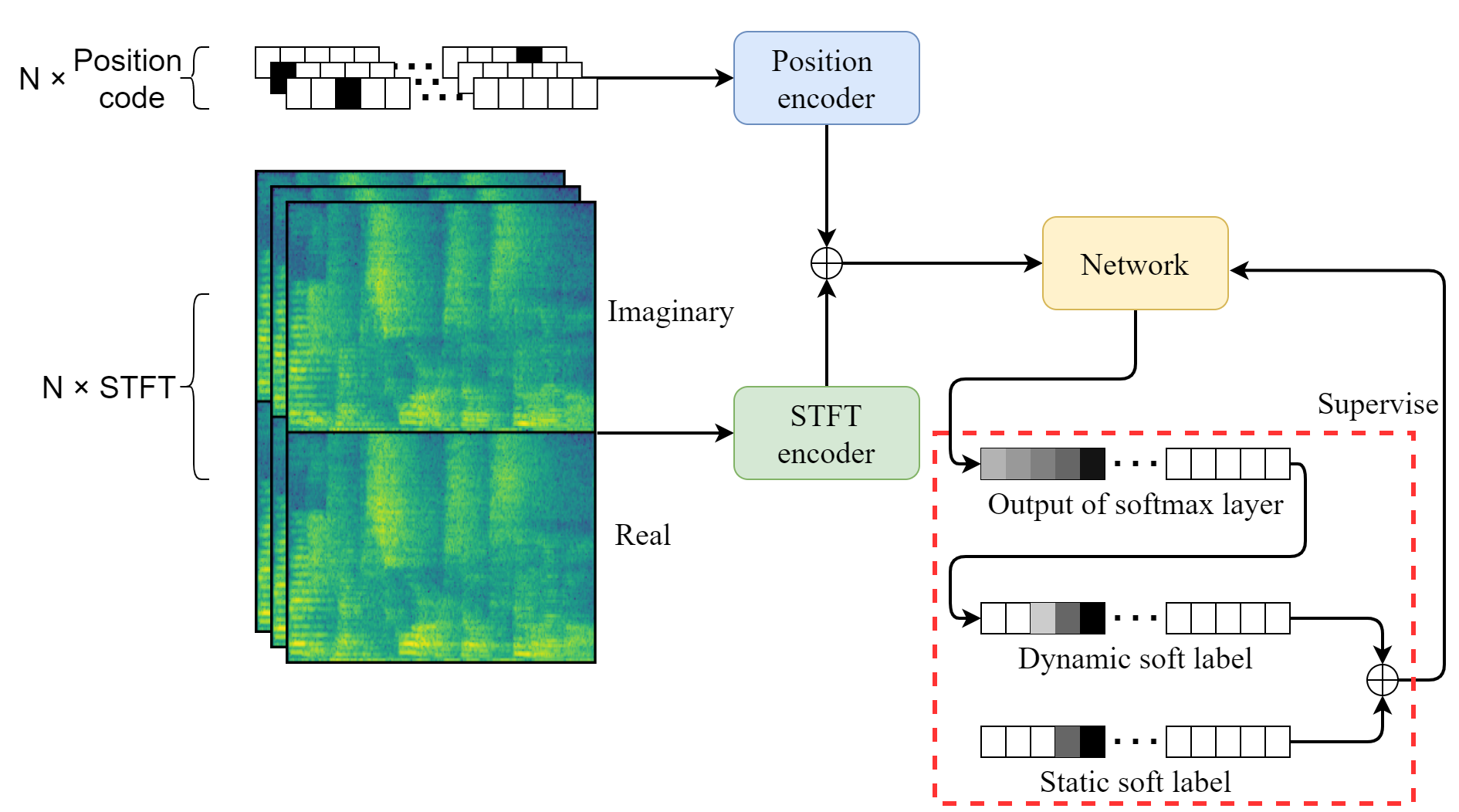}
    \caption{The proposed end-to-end sound source localization model with soft labels. The soft label generation module is highlighted in the red box.}
    \label{fig:algorithm}
\end{figure}

As shown in Fig. \ref{fig:room}, the end-to-end sound source localization with ad-hoc microphone arrays partitions a targeted room space into local areas, each of which is encoded by a one-hot code. The architecture of the method, which is shown as part of Fig. \ref{fig:algorithm} that does not include the module in the red box, takes the one-hot codes of the positions of the ad-hoc nodes as spatial input features, and takes the short-term Fourier transform (STFT) from the ad-hoc nodes as acoustic features. It formulates the sound source localization problem as a classification problem, which aims to classify a sound source into one of the grids. Because the spatial information is encoded in the one-hot codes, it could work with a special scenario where each ad-hoc node contains only a single microphone, which is the working scenario of this paper.

The original end-to-end sound source localization takes the one-hot coding to encode the room space.
Suppose the set of the one-hot codes for an $n$-class classification problem is $\mathcal{S}^{\mathrm{1-hot}} = \{\mathbf{S}^{\mathrm{1-hot}}_i\}_{i=1}^n$, with the code for the $i$th class $\mathbf{S}^{\mathrm{1-hot}}_i = [{S}^{\mathrm{1-hot}}_{i,1},\ldots,{S}^{\mathrm{1-hot}}_{i,k},$ $\ldots, {S}^{\mathrm{1-hot}}_{i,n}]$ defined as:
\begin{equation}
{S}^{\mathrm{1-hot}}_{i,k} = \left\{\begin{array}{ll}
  1,& \mbox{ if   } k = i\\
  0,& \mbox{ otherwise}
\end{array}\right.,\quad \forall k = 1,\ldots, n
\end{equation}
where $n$ is the total number of local areas.

In the following of this section, we focus on introducing the novel module highlighted in the red box in Fig. \ref{fig:algorithm}.

\subsection{Static soft label coding}

Unlike conventional classification problems where the classes may be fundamentally different and uncorrelated, the classes in our sound source localization problem are correlated. For example, we know the geometric location and similarity of the local areas, such as the distance between the centers of any two local areas. Therefore, one-hot coding may not be suitable to our problem.

Here we propose SSLC to address this issue. The SSLC for the $i$th class, denoted as $\mathbf{S}^{\mathrm{SSLC}}_i = [{S}^{\mathrm{SSLC}}_{1},\ldots,{S}^{\mathrm{SSLC}}_{i},$ $\ldots, {S}^{\mathrm{SSLC}}_{n}]$, sets its $k$th element according to the distance between the centers of the $i$th and $k$th local areas. Formally, SSLC is defined as a Gaussian function of the distance:
\begin{small}
\begin{equation}
    S_{i,k}^{\mathrm{SSLC}}=\left\{\begin{array}{ll}
    \frac{1}{\sqrt{2 \pi} \sigma} \int_{-\infty}^{-d_{i,k}} e^{-\frac{\theta^2}{2 \sigma^2}} d \theta, & \mbox{if } k\neq i\\
    {1 - {\sum_{\forall 1\le j \le n, j\neq i}^n S_{j, k}}}, & \mbox{otherwise}
    \end{array}\right.
    \forall k=1,\ldots,n
\end{equation}
\end{small}
where $d_{i,k}$ is the distance between the $i$th and $k$th local areas, $\sigma$ is the standard deviation that controls the spread of the Gaussian distribution. We set $\sigma$ = ${l_{\mathrm{ave}}}/{\alpha}_s$ where $l_{\mathrm{ave}}$ is the average diagonal length of the local areas of all rooms in the training set, and ${\alpha}_s$ is a tunable hyperparameter.

Given the above formulation, we are able to train the classification model by using SSLC, i.e. $\{\mathbf{S}_i^{\mathrm{SSLC}}\}_{i=1}^n$, as the ground-truth label instead of the one-hot coding.
The training loss of a single utterance can be:
\begin{equation}
    L^{\mathrm{SSLC}}=-\sum_{k=1}^n S_{i, k}^{\mathrm{SSLC}} \log \hat{y}_k
\end{equation}
where $\boldsymbol{x}$ is the input feature of the model, and $\hat{y}_k$ is the softmax output at the $k$th unit.

\subsection{Dynamic soft label coding}

SSLC is a handcrafted label, which may not reflect the correlation between the local areas fully. To address this issue, here we propose to learn a coding book, named DSLC, via online label smoothing \cite{zhang2021delving} in the training stage.

DSLC is obtained with the guidance of the ground-truth one-hot labels in the training stage. Specifically, for the $t$th epoch ($t \ge 1$), suppose the softmax output of the classification model for an input utterance is $\hat{\mathbf{y}}_j = [\hat{y}_1,\ldots, \hat{y}_n]$, and the ground-truth one-hot code of the speaker position of the utterance is $\mathbf{S}_{i}^{\mathrm{1-hot}}$. If the speaker position is predicted correctly, i.e. $\arg\max{\hat{\mathbf{y}}}=i$, then we pick $\hat{\mathbf{y}}$ into a set denoted as $\mathcal{Y}^{(t)}_i$; otherwise, we discard $\hat{\mathbf{y}}$.

Then, we generate a DSLC for the $i$th class by averaging the elements of $\mathcal{Y}^{(t)}_i$:
\begin{equation}
    \mathbf{S}_{i}^{{(t)}} = \frac{1}{\left|\mathcal{Y}^{(t)}_i\right|}\sum_{\forall\hat{\mathbf{y}}\in \mathcal{Y}^{(t)}_i} \hat{\mathbf{y}},\quad \forall i=1,\ldots, n
      \label{eq:dyna}
\end{equation}

Finally, for the $(t+1)$th epoch, we use DSLC generated in the $t$th epoch, i.e. $\{\mathbf{S}_{i}^{{(t)}}\}_{i=1}^n$, as the supervision to guide the training, where the training loss of of a single utterance in this epoch is defined as:
\begin{equation}
    L^{\mathrm{DSLC}}_{(t+1)}=-\sum_{k=1}^n {S}_{i,k}^{{(t)}} \log \hat{y}_k
\end{equation}
where ${S}_{i,k}^{(t)}$ is the $k$th element of $\mathbf{S}_{i}^{{(t)}}$.

Note that, before the first epoch, i.e. $t=0$, we use the vanilla label smoothing \cite{szegedy2016rethinking} to initialize $\{\mathbf{S}^{(0)}_i\}_{i=1}^n$.

\subsection{Joint training}

The reliability of DSLC at the $(t+1)$th epoch depends on the deep model at the $t$th epoch. Because the initial deep model is too weak, training DSLC from the beginning is very difficult and sometimes unconverged.

To address this issue, we propose a joint training strategy, which uses static labels, such as SSLC or one-hot codes, to assist the training of DSLC:
\begin{equation}
    L^{\mathrm{joint}}_{(t)} = \alpha_d^{(t)} L^{\mathrm{DSLC}}_{(t)} + (1-\alpha_d^{(t)})L^{\mathrm{SSLC}}
\end{equation}
where $\alpha_d^{(t)}$ is a tunable hyperparameter at the $t$th epoch.

This raises a question of how to set $\alpha_d^{(t)}$. The first choice is to set $\alpha_d^{(1)}=\alpha_d^{(2)}=\ldots=\alpha_d^{(T)}$ to a constant $\alpha_d$. Another choice is to adaptively adjust $\alpha_d^{(t)}$ according to the importance of $L^{\mathrm{DSLC}}_{(t)}$. In this paper, we set $\alpha_d^{(t)} = \mathrm{ACC}^{(t)}$ , where $\mathrm{ACC}^{(t)}$ is the prediction accuracy of the $t$th model on the training set. Of course, we also can set $\alpha_d^{(t)}$ a learnable parameter of the deep model.

\section{Experiments}
\label{sec:experiments}

\subsection{Experimental setup}
We conducted experiments on simulated data. The source speech is from the LibriSpeech corpus \cite{panayotov2015librispeech}. The train-clean-360 subset, dev-clean subset and test-clean subset of the corpus contain 921, 40 and 40 speakers respectively. We used voice activity detection (VAD) provided by the Torchaudio module \cite{yang2022torchaudio} to remove silent segments. The speech segments that are less than 2 seconds after VAD were discarded. For each of the remaining segments, we randomly selected a 2-second piece to generate multi-channel data with reverberation. We used the Pyroomacoustics \cite{scheibler2018pyroomacoustics} module to generate room impulse responses. The reverberation time T$_{60}$ of all simulated speech follows a uniform distribution ranging from $[0.3, 1.0]$ second. We further added additive noise to the reverberant speech. The additive noise was randomly selected from a {large scale noise set} that contains 126 hours of different types of noise segments. The signal-to-noise ratio of an utterance was selected randomly from a range of $[20, 50]$ dB. Eventually, the number of utterances in the training, validation, and test sets are $18000$, $2500$, and $2500$, respectively.

We generated 15 rooms of different sizes, where 10 rooms were used for training, and 5 rooms were used for testing.
 The length and width of the rooms range from 4m to 10m. The height of the rooms was fixed at 3m. The height of the microphone arrays and sound source was fixed at 1m. Each room was divided into $15\times15$ local areas in 2D. For each utterance, we first randomly picked a room, then randomly set its speech source in one of the 225 local areas, and finally randomly placed 30 ad-hoc microphone nodes in the other 224 local areas.

We used the DNN model in \cite{gong2022e2e} as our backbone network.
First, it uses the ad-hoc node's position code and STFT acoustic feature as the inputs for two encoders, respectively. The concatenated outputs of the two encoders are fed into a spatial-temporal attention network in the following phase. The attention network sequentially performs cross-channel attention, channel fusion, and cross-frame attention. The local region of the speaker is considered to be the average output along the time axis of the spatial-temporal attention network.
ReLU was used as the activation function, and Adam was used as the optimizer to train the model.


\begin{table*}[t]
  \centering
  \caption{MAE (in meters) and ACC (\%) results of the proposed five soft label coding methods and the one-hot coding.}
  \label{tab:simu_res}
  \vspace{6pt}
  \scalebox{0.8}{
    \begin{tabular}{lcccccccccccc}
    \toprule
    \multicolumn{1}{c}{\multirow{2}[4]{*}{\textbf{Method}}} & \multicolumn{2}{c}{\textbf{Room1}} & \multicolumn{2}{c}{\textbf{Room2}} & \multicolumn{2}{c}{\textbf{Room3}} & \multicolumn{2}{c}{\textbf{Room4}} & \multicolumn{2}{c}{\textbf{Room5}} & \multicolumn{2}{c}{\textbf{Average}} \\
\cmidrule{2-13}          & MAE & ACC & MAE & ACC & MAE  & ACC  & MAE  & ACC  & MAE  & ACC  & MAE  & ACC  \\
    \midrule
    One hot \cite{gong2022e2e} & 0.2135 & 54.6  & 0.2471 & 51.0    & 0.2959 & 46.6  & 0.3283 & \textbf{49.8} & 0.3207 & 54.0  & 0.2811 & 51.2 \\

    SSLC   & 0.2260 & 53.4  & 0.2566 & 49.2  & 0.3001 & \textbf{48.8}  & 0.3363 & 43.6  & 0.3344 & 52.8  & 0.2907 & 49.6 \\

    DSLC+One hot ($\alpha_d$ is a constant) & 0.2226 & 54.2  & 0.2592 & 47.6  & 0.3079 & 43.0    & 0.3234 & 47.4  & 0.3225 & 55.4  & 0.2871 & 49.5 \\

    \textbf{DSLC+SSLC ($\alpha_d$ is a constant)} & 0.2115 & \textbf{57.6} & \textbf{0.2448} & \textbf{52.4} & \textbf{0.2873} & 47.0    & 0.3088 & 48.2  & \textbf{0.3185} & \textbf{56.0} & \textbf{0.2742} & \textbf{52.2} \\

    DSLC+One hot ($\alpha_d^{(t)}$ is a variable) & 0.2149 & 56.0    & 0.2689 & 45.0    & 0.3052 & 45.0    & 0.3239 & 49.6  & 0.3408 & 47.8  & 0.2907 & 48.7 \\

    DSLC+SSLC ($\alpha_d^{(t)}$ is a variable) & \textbf{0.2059} & 57.2  & 0.2577 & 47.8  & 0.2974 & 47.8 & \textbf{0.2998} & 47.4  & 0.3351 & 53.2  & 0.2792 & 50.7 \\
    \bottomrule
    \end{tabular}}
\end{table*}%

We used mean absolute error (\textbf{MAE}) to evaluate the proposed method:
\begin{equation}
    \mathrm{MAE}=\frac{1}{I} \sum_{i=1}^I \sqrt{\left(x_i^{spkr}-\tilde{x}_i^{spkr}\right)^2
    + \left(y_i^{spkr}-\tilde{y}_i^{spkr}\right)^2}
\end{equation}
where $I$ is the number of the test utterances, $(x_i^{spkr}, y_i^{spkr})$ is the ground truth coordinate of the $i$th speaker position, and $(\tilde{x}_i^{spkr}, \tilde{y}_i^{spkr})$ is the coordinate of the center of the local area of the predicted speaker position $\tilde{y}_i$.

Because the region in a local area is quantized to the center of the local area, the \textbf{quantization error} inevitably exists as an upperbound of any methods based on the classification formulation of the sound source localization. We denote the upperbound as UB-MAE.
In this paper, the average UB-MAE over the 5 test rooms is 0.1693m.

\subsection{Results}

We trained sound source localization models with 6 label coding methods, as summarized in Table \ref{tab:simu_res}. From the results, we see that the proposed joint training method {DSLC+SSLC ($\alpha_d$ is constant)} generally outperforms one-hot. In most cases, the higher the ACC, the smaller the MAE. The performance of the joint training is always better than the one-hot coding, no matter whether $\alpha_d^{(t)}$ is a constant or a variable. The schemes of adaptively adjusting $\alpha_d$ does not achieve better results than the schemes of setting $\alpha_d$ to a constant. This is probably caused by that setting $\alpha_d^{(t)}=\mathrm{ACC}^{(t)}$ makes DSLC overfit to the training set.

\begin{table}[t]
  \centering
  \caption{Learning error (in meters) comparison betwen one-hot coding and the proposed {DSLC+SSLC ($\alpha_d$ is a constant)} method. }
  \scalebox{0.8}{
    \begin{tabular}{cccc}
    \toprule
    Room & One hot & DSLC+SSLC & Relative reduction (\%) \\
    \midrule
    Room1   & 0.0701 & 0.0681 & 2.83 \\
    Room2   & 0.0942 & 0.0919 & 2.44 \\
    Room3   & 0.1310 & 0.1224 & 6.56 \\
    Room4   & 0.1484 & 0.1289 & 13.14 \\
    Room5   & 0.1156 & 0.1134 & 1.90 \\
    Average & 0.1119 & 0.1049 & 6.26 \\
    \bottomrule
    \end{tabular}}%
  \label{tab:gap}%
\end{table}%

Because the quantization error always exist, the error in Table \ref{tab:simu_res} contains two parts, one from the quantization error, and the other one from the \textit{learning error} caused by the coding methods.
{We further investigated the learning error independently in Table \ref{tab:gap} and found our proposed method achieved a relative 6.26\% average reduction.}

\begin{table}[t]
  \centering
  \caption{Effect of ${\alpha}_s$ in SSLC on performance.}
    \scalebox{0.85}{
    \begin{tabular}{lrrrrr}
    \toprule
    ${\alpha}_s$     & 2.6   & 2.7   & 2.8   & 2.9   & 3 \\
    \midrule
    MAE (m) & 0.2793 & 0.2796 & 0.2778 & 0.2782 & 0.2822 \\
    ACC (\%) & 50.5  & 51.7  & 51.0    & 51.6  & 50.7 \\
    \bottomrule
    \end{tabular}%
    }
  \label{tab:ssl}%
\end{table}%

There are two tunable hyperparameters, ${\alpha}_s$ in SSLC and $\alpha_d$ in ``{DSLC+SSLC ($\alpha_d$ is a constant)}''. Here we study the their effects on performance. For ${\alpha}_s$ in SSLC, when ${\alpha}_s<2.6$, the probability that the speaker falls in the ground-truth local region is lower than 0.8
which results in poor performance; when $\alpha_s>3.0$, SSLC approaches very close to one-hot coding. Therefore, we studied $\alpha_s\in [2.6, 3.0]$ in Table \ref{tab:ssl}. From the table, we see that the optimal value of ${\alpha}_s$ is 2.8. For ${\alpha}_d$ in ``{DSLC+SSLC ($\alpha_d$ is a constant)}'', we searched ${\alpha}_d$ from a range of $(0,1)$, and find that, as shown in Table \ref{tab:dsl}, the optimal value of ${\alpha}_d$ is 0.5.


\begin{table}[t]
  \centering
  \caption{Effect of $\alpha_d$ in ``{DSLC+SSLC ($\alpha_d$ is a constant)}'' on performance.}
    \scalebox{0.85}{
    \begin{tabular}{lrrrrr}
    \toprule
    $\alpha_d$     & 0.2   & 0.4   & 0.5   & 0.6   & 0.8 \\
    \midrule
    MAE (m) & 0.2839 & 0.2741 & 0.2726 & 0.2772 & 0.2845 \\
    ACC (\%) & 49.2  & 50.6  & 51.4  & 49.6  & 50.2 \\
    \bottomrule
    \end{tabular}%
    }
  \label{tab:dsl}%
\end{table}%

\section{Conclusion}
\label{sec:conclusion}

In this paper, we propose a label smoothing algorithm against the weakness of the one-hot coding in the end-to-end sound source localization with ad-hoc microphone arrays.
{It includes a handcrafted SSLC, calculated from local area center distances, and a learnable DSLC, updated during training process using correctly classified sample statistics.}
Experimental results show the effectiveness of the proposed algorithm over the one-hot coding, without increasing the model complexity and training time.

\vfill\pagebreak
\bibliographystyle{IEEEbib}
\bibliography{refs}

\end{document}